# Machine Learning Approaches for Modeling Spammer Behavior

Md. Saiful Islam[1], Abdullah Al Mahmud[2] and Md. Rafiqul Islam[3]

[1]Institute of Information Technology, University of Dhaka, Dhaka 1000, Bangladesh
[2]Dept. of Computer Science and Engineering, Ahsanullah University of Science and Technology, Bangladesh
[3]School of Information Technology, Deakin University, Melbourne, VIC 3216, Australia
saifulit@univdhaka.edu, aamrubel@gmail.com and rislam@deakin.edu.au

**Abstract.** Spam is commonly known as unsolicited or unwanted email messages in the Internet causing potential threat to Internet Security. Users spend a valuable amount of time deleting spam emails. More importantly, ever increasing spam emails occupy server storage space and consume network bandwidth. Keyword-based spam email filtering strategies will eventually be less successful to model spammer behavior as the spammer constantly changes their tricks to circumvent these filters. The evasive tactics that the spammer uses are patterns and these patterns can be modeled to combat spam. This paper investigates the possibilities of modeling spammer behavioral patterns by well-known classification algorithms such as Naïve Bayesian classifier (Naïve Bayes), Decision Tree Induction (DTI) and Support Vector Machines (SVMs). Preliminary experimental results demonstrate a promising detection rate of around 92%, which is considerably an enhancement of performance compared to similar spammer behavior modeling research.

**Keywords:** Spam Email, MLA, Naïve Bayes, DTI, SVMs.

## 1   Introduction

The exponential growth of spam emails in recent years is a fact of life. Internet subscribers world-wide are unwittingly paying an estimated €10 billion a year in connection costs just to receive "junk" emails, according to a study undertaken for the European Commission [1]. Though there is no universal definition of spam, unwanted and unsolicited commercial email is basically known as the junk email or spam to the internet community. Spam's direct effects include the consumption of computer and network resources and the cost in human time and attention of dismissing unwanted messages [2].

   Combating spam is a difficult job contrast to the spamming. The simplest and most common approaches are to use filters that screen messages based upon the presence of common words or phrases common to junk e-mail. Other simplistic approaches include *blacklisting* (automatic rejection of messages received from the addresses of known spammers) and *whitelisting* (automatic acceptance of message received from known and trusted correspondents). The major flaw in the first two approaches is that it relies upon complacence by the spammers by assuming that they are not likely to change (or forge) their identities or to alter the style and vocabulary of their sales pitches. Whitelisting risks the possibility that the recipient will miss legitimate e-mail from a known or expected correspondent with a heretofore unknown address, such as correspondence from a long-lost friend, or a purchase confirmation pertaining to a transaction with an online retailer. A detail explanation of these techniques is given in [3].

Machine learning algorithms namely Naïve Bayesian classifier, Decision Tree induction and Support Vector Machines based on keywords or tokens extracted from the email's *Subject*, *Content-Type* Header and Message *Body* have been used successfully in the past [2],[3],[4],[5]. Very soon they fall short to filter out spam emails as the spammer changing themselves in the ways that are very difficult to model by simple keywords or tokens [6].

The tactics the spammer uses follows patterns and these behavioral patterns can be modeled to combat spam. Actually the more they try to hide, the easier it is to see them [6]. This study investigates the possibilities of modeling spammer behavioral patterns instead of vocabulary as features for spam email categorization. The three well-known machine learning algorithms Naïve Bayes, DTI and SVMs are experimented to model common spammer patterns, as these classifiers has already shown great performance in different research in spam classifier [3], [4], [5]. Among the classifiers, Naïve Bayes shows its best suitability.

The paper is organized as follows: section 2 discusses the three machine learning algorithms (MLAs); section 3 presents common spammer patterns, email corpus, feature construction and evaluation measures; section 4 discusses the experimental results; and finally section 5 concludes the paper.

## 2 Machine Learning Algorithms

The success of machine learning algorithms in text categorization (TC) has led researchers to investigate learning algorithms for filtering spam emails [3], [4], [5]. This paper studies the following three machine learning algorithms to model spammer tricks and techniques.

### 2.1 Naïve Bayesian Classifier

Bayesian classifiers are based on Bayes' theorem. For a training e-mail E, the classifier calculates for each category, the probability that the e-mail should be classified under $C_i$, where $C_i$ is the $i^{th}$ category, making use of the law of the conditional probability:

$$P(C_i|E) = \frac{P(C_i)P(E|C_i)}{P(E)}$$

Assuming class conditional independence, that is, the probability of each word in an e-mail is independent of the word's context and its position in the e-mail, $P(E|C_i)$ can be calculated as the product of each individual word $w_j$'s probabilities appearing in the category $C_i$ ($w_j$ being the $j^{th}$ of $l$ words in the e-mail):

$$P(E|C_i) = \prod_{j=1}^{l} P(w_j|C_i)$$

The category maximizing $P(C_i | E)$ is predicted by the classifier [5], [7].

### 2.2 Decision Tree Induction

A decision tree is a *flow-chart-like* tree structure, where each internal node denotes a test on an attribute, each branch represents an output of the test, and leaf nodes correspond to the classification result [5], [7]. The topmost node in the tree is the root node. An example of a typical decision tree is given below:

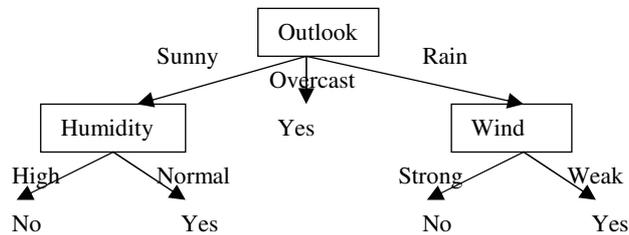

**Fig. 1.** A decision tree used to determine whether it is suitable to play tennis.

To classify an unknown sample the attribute values of the sample are tested against the tree and a path will be traced starting the root to a leaf node that identifies the class prediction for that sample. The commonly used rule learning algorithms J48, ID3 and C4.5 are based on decision trees [5]. The advantage offered by the decision trees is that it can easily be converted to decision rules and comprehended even by a naïve user [7].

### 2.3 Support Vector Machines

Support vector machines (SVM) are a collection of supervised learning methods that can be applied to classification or regression [4], [7], [8]. Viewing input data as two sets of vectors in a *d*-dimensional space, an SVM constructs a separating *hyperplane* in that space, one which maximizes the *margin* between the two data sets.

Suppose we are given some training data, a set of points of the form:

$$D = \{(x_i, c_i) \mid x_i \in \Re^d, c_i \in \{-1, 1\}\}_{i=1}^{n}$$

where the $c_i$ is either 1 or −1, indicating the class to which the point belongs. Each $x_i$ is a *d*-dimensional real vector. We want to find the maximum-margin hyperplane which divides the points having $c_i = 1$ from those having $c_i = -1$. Any hyperplane can be written as the set of points $X$ satisfying

$$W \bullet X - b = 0$$

where $\bullet$ denotes the dot product. The vector $W$ is a normal vector: it is perpendicular to the hyperplane. The parameter $\dfrac{b}{\|W\|}$ determines the offset of the hyperplane from the origin along the normal vector $W$.

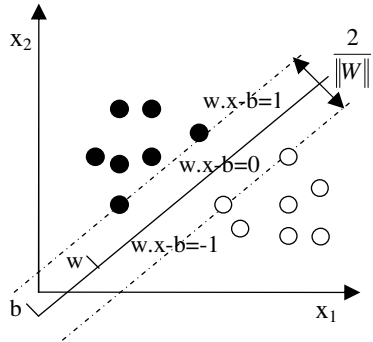

**Fig. 2**. Maximum-margin hyperplane and margins for a SVM trained with samples from two classes. Samples on the margin are called the support vectors [7], [8].

We want to choose the $W$ and b to maximize the margin, or distance between the parallel hyperplanes that are as far apart as possible while still separating the data. These hyperplanes can be described by the equations:

$$W \bullet X - b = 1$$

and

$$W \bullet X - b = -1$$

Note that if the training data are linearly separable, we can select the two hyperplanes of the margin in a way that there are no points between them and then try to maximize their instance. By using geometry, we find the distance between these two hyperplanes is $\frac{2}{\|W\|}$, so we want to minimize $\|W\|$.

As we also have to prevent data points falling into the margin, we add the following constraint: for each i either

$W \bullet x_i - b \geq 1$ for $x_i$ of the first class

or

$W \bullet x_i - b \leq -1$ for $x_i$ of the second.

We can put this together to get the optimization problem:

Minimize (in $W, b$) $\|W\|$ subject to (for any $i = 1,...,n$)

$$c_i(W \bullet x_i - b) \geq 1$$

## 3 Proposed Modeling Approach

People who create spams are called spammers. Electronic mails (emails) are the most common playground of many spammers in the Internet. A tremendous effort has already been invested by the researchers on anti-spamming techniques [4], [5], [12], [13].

### 3.1 Spammer Behavioral Patterns

The keyword-based statistical analyzers mostly depend on tokenization of the email content and extracting feature from tokenized keywords to model spammer behavior. Tokenization can be misguided in many several ways as today's email supports character sets other than ASCII, non-text attachments and bodies with multiple parts. For example, the following HTML tricks can be used to do this:

GET<!-- banana -->V<!-- 45-->I<!-- wumpus -->A<!-- dskfj -->G <!-- adf -->R<!-- free -->A

Thus above nonsense HTML tags only split the special word "viagra" and disguise the tokenizer though it would be shown as "GET VIAGRA" to email client.

Even a word can be replaced with characters of other languages or like same character. For example, "V1DEO" can be send instead of "VIDEO" and "Fántástìç" instead of "Fantastic". A combination of special characters can used to produce alphabetical characters. For example, char "V" can be represented as the combination of right slash"\" and left slash "/". A list of these kinds of techniques can be found in [9]. A grouping or clustering of these techniques is given Table 1 for quick review.

Table 1 has 30 different tricks and one can easily verify that HTML based tactics cover most of them (70%). It can also be shown that 75% of Cascading Style Sheet

(CSS) and 50% of Image-based tricks are also covered by HTML-based tactics. It is evident from table 1 that Java Script and MIME (and/or others) based tricks do not overlap with HTML/CSS based tactics.

**Table 1.** Common spammer tricks.

|  | Java Script | Image | CSS | HTML | MIME/Others |
|---|---|---|---|---|---|
| Title Case |  |  |  | Y |  |
| Sticky Finger |  |  |  | Y |  |
| Accent |  |  |  |  | Y |
| Readable Spell |  |  |  | Y |  |
| Dot Matrix |  |  | Y | Y |  |
| Right-to-Left |  |  |  | Y |  |
| HTML Numbers |  |  |  | Y |  |
| Comments |  |  |  | Y |  |
| Styles |  |  | Y |  |  |
| Invisible Ink |  |  | Y | Y |  |
| Matrix |  |  | Y | Y |  |
| Encoding of MSG |  |  |  |  | Y |
| Encrypted Message Bodies | Y |  |  |  |  |
| Copperfield |  |  | Y |  |  |
| Invisible Image |  | Y |  |  |  |
| Zero Image |  | Y | Y | Y |  |
| Slice and Dice |  | Y | Y | Y |  |
| Cross Word |  |  | Y | Y |  |
| Honorary Title |  |  |  | Y |  |
| Image Chopping |  | Y |  |  |  |
| Cramp |  |  |  | Y |  |
| Framed |  |  |  | Y |  |
| Big Tag |  |  |  | Y |  |
| Fake Text |  |  |  | Y |  |
| Slick Click |  |  |  | Y |  |
| Phishing |  |  |  | Y |  |
| False Click |  |  |  | Y |  |
| Pump & Dump |  |  |  |  | Y |
| I'm Feeling Lucky |  |  |  | Y |  |

In this study, a model has been developed exploiting machine learning algorithms to capture common spammer patterns instead of keyword analysis. The 21 handy crafted features from each e-mail message extracted from subject header, priority & content-type headers and body shown in Table 2 simulate all possible common spammer tricks. These features have also been optimized in their capability of classifying spam emails. The rationale of these features can be verified by their statistics both in spam and non-spam emails. For example, whether a content-type header appeared within the message headers or whether the content type had been set to "text/html" is a common feature of spam, as our investigation revealed. The corpus that has been used in our experimentation, we observed that 98% spam emails include this feature. Similarly, color element (both CSS and HTML format) is also a frequent feature of spam emails. Colorful images those are generally included in the email for X-rated and unwanted internet marketing groups send to catch users' attention. The use of color elements in non-spam mails is very low. We found that that 56% spam emails contain color elements whereas it exists only for 10% non-spam emails. The inclusion of this feature in our classification has improved performance considerably, which shows its practicality. We also added feature 19-21 as in Table 2, which are significant features of recent spams.

**Table 2.** Features extracted from each e-mail.

| Feature | Category 1: Features From the Message Subject Header |
|---|---|
| 1 | Binary feature indicating 3 or more repeated characters |
| 2 | Number of words with all letters in uppercase |
| 3 | Number of words with at least 15 characters |
| 4 | Number of words with at least two of letters J, K, Q, X, Z |
| 5 | Number of words with no vowels |
| 6 | Number of words with non-English characters, special characters such as punctuation, or digits at beginning or middle of word |
| | **Category 2: Features From the Priority and Content-Type Headers** |
| 7 | Binary feature indicating whether the priority had been set to any level besides normal or medium |
| 8 | Binary feature indicating whether a content-type header appeared within the message headers or whether the content type had been set to "text/html" |
| | **Category 3: Features From the Message Body** |
| 9 | Proportion of alphabetic words with no vowels and at least 7 characters |
| 10 | Proportion of alphabetic words with at least two of letters J, K, Q, X, Z |
| 11 | Proportion of alphabetic words at least 15 characters long |
| 12 | Binary feature indicating whether the strings "From:" and "To:" were both present |
| 13 | Number of HTML opening comment tags |
| 14 | Number of hyperlinks ("href=") |
| 15 | Number of clickable images represented in HTML |
| 16 | Binary feature indicating whether a text color was set to white |
| 17 | Number of URLs in hyperlinks with digits or "&", "%", or "@" |
| 18 | Number of color element (both CSS and HTML format) |
| 19 | Binary feature indicating whether JavaScript has been used or not |
| 20 | Binary feature indicating whether CSS has been used or not |
| 21 | Binary feature indicating opening tag of table |

### 3.2 Email Corpus

Classification based spam filtering systems have two major drawbacks. Firstly, building a perfect data set free from noise or imperfection as noise adversely affect the classifier's performance [12]. The nature of spam email is very dynamic and the content of email is textually misleading due to obfuscation as we explained earlier. This remains a continuous challenge for spam filtering techniques. Secondly, most training models of the classifier have limitations on their operations [14]. Classifiers often produce uncorrelated training errors due to the dimension of feature space; a dissimilar output space is generated for changing feature space from small dimension to complex high dimension.

In this work a corpus of 1,000 emails received over a period of several months is used for experimentation. The distribution of both spam and non-spam emails in this collection is equal. The equal distribution is preferred to make the classifier to eliminate the biasness towards a particular category. That is, out of 1,000 emails 500 is spam and 500 is non-spam. The collection of this corpus is selected over a time and latest trend in spamming is kept in mind. Also the author's experience with spam research and statistical selection methodology is applied to the selection, which made this email bank very much representative of current spamming.

### 3.3 Feature Construction

Each email is parsed as text file to identify each header element to distinguish them from the body of the message. Every substring within the subject header and the message body that was delimited by white space was considered to be a *token*, and an *alphabetic word* was defined as a token delimited by white space that contains only English alphabetic characters (A-Z, a-z)or apostrophes. The tokens were evaluated to create a set of 21 hand-crafted features from each e-mail message (Table 2) of which features 1-17 are proposed in [6]. In addition of these 17 features this study proposes other four features 18-21. The study investigates the suitability of these 21 features in classifying spam emails.

### 3.4 Evaluation Metrics

Estimating classifier accuracy is important since it allows one to evaluate how accurately a given classifier will classify unknown samples on which the classifier has not been trained. The effectiveness of a classifier is usually measured in terms of accuracy, precision and recall [5], [7]. These measures are calculated using the confusion matrix given below:

**Table 3.** Confusion matrix.

| Category $C_i$ | Correct | |
|---|---|---|
| Predicted ↓ | YES | NO |
| YES | $TP_i$ | $FP_i$ |
| NO | $FN_i$ | $TN_i$ |

TP=true positives
FP=false positives
FN=false negatives
TN=true negatives

Accuracy of a classifier is calculated by dividing the number of correctly classified samples by the total number of test samples and is defined as:

$$Accuracy = \frac{number\ of\ correctly\ classified\ samples}{total\ number\ of\ test\ samples}$$

$$= \frac{TP + TN}{TP + FP + FN + TN}$$

Precision measures the system's ability to present only relevant items while recall measures system's ability to present all relevant items. These two measures are widely used in TREC evaluation of document retrieval [10]. Precision is calculated by dividing the number of samples that are true positives by the total number of samples classified as positives and is defined as:

$$\text{Precision} = \frac{number\ of\ true\ positives}{total\ number\ of\ samples\ classified\ as\ positives}$$

$$= \frac{TP}{TP+FP}$$

Analogously, recall is calculated by dividing the number of samples that are true positives by the total number of samples that classifier should classified as positives and is defined as:

$$\text{Recall} = \frac{number\ of\ true\ positives}{total\ number\ of\ positive\ samples}$$

$$= \frac{TP}{TP+FN}$$

In this study, both precision and recall are kept close to give equal importance on both of them.

The block diagram of the proposed model of spam email classification process exploiting spammer behavioral patterns given in Fig 3.

## 4   Experimental Results and Discussion

Table 4 summarizes the comparative results of the three well-known machine learning algorithms namely Naïve Bayesian classifier, decision tree induction and SVM. These algorithms are tested on Weka 3.6.0 suite of machine learning software written in Java, developed at the University of Waikato [11]. It is observed that Naïve Bayesian classifier outperforms than other two machine learning algorithms in all cases. The highest level of accuracy that can be achieved by Naïve Bayesian classifier is 92.2% (shown in yellow color in Table 4) using features from category 2 and 3. The accuracy that can be achieved by any learning algorithms using features from category 1 is negligible. Features from category 2 and 3 contribute mostly in classifying spam emails from non-spam emails for all machine learning algorithm experimented in this study.

Highest number of features is always desirable only if their inclusion increase classifier's accuracy significantly. Growing number of features not only hinders multidimensional indexing but also increases overall execution time. So, this study starves to find an optimal number of features that can be effectively used to lean a classifier without degrading the level of accuracy.

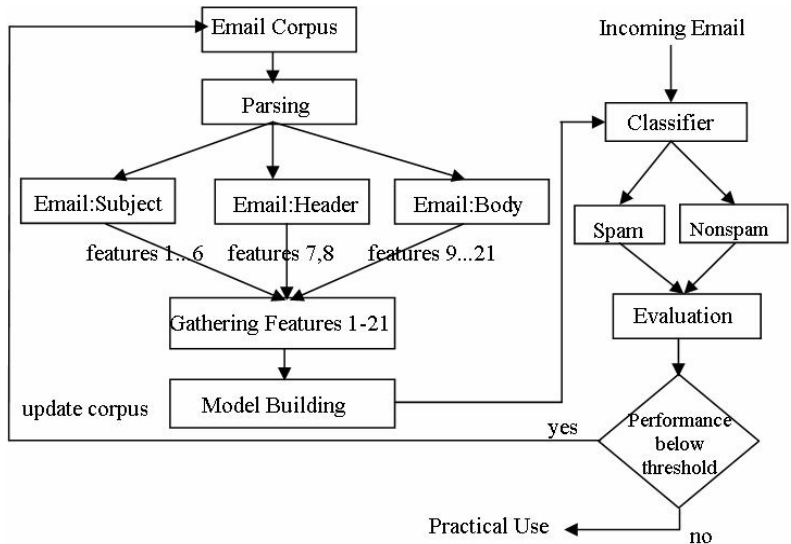

**Fig. 3.** Block diagram of the proposed model of spam email classification process.

**Table 4.** Comparison results for Naïve Bayesian Classifier, Decision Tree Induction and SVM.

| Features | Naïve Bayesian Classifier (Naïve Bayes) | | | Decision Tree Induction (J48) | | | SVM (SMO) | | |
|---|---|---|---|---|---|---|---|---|---|
| | Acc. | Pre. | Rec. | Acc. | Pre. | Rec. | Acc. | Pre. | Rec. |
| Cat1 Only | 56.5% | 55.7% | 56.5% | 67.8 % | 68.9% | 67.8% | 62.6% | 67.8% | 62.6% |
| Cat2 Only | 65.2% | 75.0% | 65.2% | 65.2 % | 75.0% | 65.2% | 65.2% | 75.0% | 65.2% |
| Cat3 Only | 88.7% | 88.7% | 88.7% | 86.9 % | 87.4% | 87.0% | 72.2% | 72.5% | 72.2% |
| Cat1+Cat2 | 66.9% | 67.3% | 67.0% | 73.9% | 74.3% | 73.9% | 68.7 | 72.6% | 68.7% |
| **Cat2+Cat3** | **92.2%** | **92.2%** | **92.2%** | 86.9 % | 87.0% | 87.0% | 82.6% | 83.7% | 80.9% |
| Cat1+Cat3 | 80.8% | 80.9% | 80.9% | 80.8 % | 80.8% | 80.9% | 76.5% | 76.8% | 76.5% |
| Cat1+ Cat2 + Cat3 | 86.9% | 87.0% | 87.0% | 84.3 % | 84.3% | 84.3% | 74.8% | 74.7% | 74.8% |

Applying best first forward attribute selection method the study gets only 10 features from category 2 and category 3 useful for classifying the spam and non-spam emails without sacrificing the accuracy as shown in Table 5. The set includes features 8, 9, 10, 12, 13, 14, 15, 16, 17, and 18 of which feature 18 is identified in this study. The Naïve Bayesian classifier again outperforms other two learning algorithms. The optimal feature set obtained by applying best first forward attribute selection method for the features proposed in [6] includes only features 8, 9, 10, 12, 13, 14, 15, 16 and 17, a total of 9 features. In this case decision tree induction outperforms other two machine learning algorithms (shown in light blue in Table 5).

**Table 5.** Evaluation of learning algorithms with optimal feature set.

| Features | Naïve Bayesian Classifier(Naïve Bayes) | | | Decision Tree Induction (J48) | | | SVM (SMO) | | |
|---|---|---|---|---|---|---|---|---|---|
| | Acc. | Pre. | Rec. | Acc. | Pre. | Rec. | Acc. | Pre. | Rec. |
| $F_1$ | 92.2% | 92.2% | 92.2% | 89.6% | 89.9% | 89.6% | 83.5% | 85.5% | 83.5% |
| $F_2$ | 86.1% | 87.4% | 86.1% | 91.3% | 91.3% | 91.3% | 83.5% | 85.5% | 83.5% |

*$F_1$: {8, 9, 10, 12, 13, 14, 15, 16, 17, 18} - identified by this study
**$F_2$: {8, 9, 10, 12, 13, 14, 15, 16, 17} - identified in [**6**]

Though the study presented in [6] uses neural network for modeling spammer common patterns and achieved similar performance, but the limitation of neural network is its longer training time and inherent complexity of explaining its derivation, degraded the approach. On the contrary, Bayesian Classifier has the advantage of incremental inclusion of features and beforehand calculation. The decision tree based classification offers the best expressive power and allow better understanding about the classification process and knowledge adoption. Therefore, the proposed modeling approach will have added advantage in this regard.

## 5 Conclusion and Future Work

This paper studies the modeling of spammer behavior by the well-known machine learning algorithms for spam email classification. Based on examining different features and different learning algorithms, the following conclusions can be drawn from the study presented in this paper:

- Spammer behavior can be modeled using features extracted from Content-Type header and message Body only.
- The contribution of features extracted from subject header in spam email detection is negligible or insignificant.
- Naïve Bayesian classifier best models the spammer behavior than other two machine learning techniques namely Decision Tree Induction and SVMs.
- It is possible to get an optimal number of features that can be effectively applied to learning algorithms to classify spam emails without sacrificing accuracy.

The preliminary result presented in this study seems promising in modeling spammer common behavioral patterns compared to similar research. As Naïve Bayes and DTI both offers cost effective framework in classifying spam emails [3], we are focusing our experiment with established spam data and benchmarks. Naïve Bayes has the advantage of incremental inclusion and/or exclusion of features and DTI offers best expressive power. So, natural progression will be combining these two ML algorithms in multi-core architecture [13], running both classifier simultaneously in different cores to minimize time and applying voting mechanism to increase positivity, which will give best opportunity to model spammer common patterns. We are also working on developing multi-classifier based spam filters [14] exploiting spammer behavioral patterns.

The contribution of this paper is threefold: it shows why keyword based spam email classifier may fail to model spammers' altering tricks, common patterns adopted by spammers and the rationale of using these patterns against them to combat spam; suitability of modeling spammer common patterns using machine learning algorithms and finally, establishment of the four concluding remarks.

# References

1. Data protection: "Junk" E-mail Costs Internet Users 10 Billion a Year Worldwide - Commission Study. In: http://europa.eu/rapid/pressReleasesAction.do?reference=IP/01/154, last accessed on 14$^{th}$ Feb, 2009.
2. Aery, M., Chakravarthy, S.: eMailSift: Email Classification Based on Structure and Content. In: Proc. of 5$^{th}$ IEEE Intl. Conf. on Data Mining, pp. 1-8 (2005)
3. Islam, M. R, Chowdhury, M. U.: Spam Filtering using ML Algorithms. In: Proc. of IADIS International Conf. on WWW/Internet, pp. 419-426 (2005)
4. Drucker, H., Wu, D., Vapnik, V. N.: Support Vector Machines for Spam Categorization", IEEE Transactions on Neural Networks, Vol. 10, No. 5, pp. 1048-1054, 1999
5. Eichler, K.: Automatic Classification of Swedish Email Messages. B.A Thesis, Eberhard-Karls-Universitat Tubingen (2005)
6. Stuart, J. I., Cha, S., Tappert, C.: A Neural Network Classifier for Junk E-mail. In: Lecture notes in computer science, Vol. 3163, pp. 442-450 (2004)
7. Han, J., Kamber, M.: Data Mining Concepts and Techniques. Academic Press, ISBN 81-7867-023-2 (2001)
8. Islam, M. S., Amin, M.I.: An Architecture of Active Learning SVMs with Relevance Feedback for Classifying E-mail. In: Journal of Computer Science, Vol. 1, No. 1, pp. 15-18 (2007)
9. Common Spammer Techniques. www.process.com/precisemail/spamtricks.pdf
10. Makhoul, J., Kubala, F., Schwartz, R., Weischedel, R.: Performance Measures for Information Extraction. In: Proc. of DARPA Broadcast News Workshop, Herndon, VA (1999)
11. Holmes, G., Donkin, A., Witten, I. H.: Weka: A machine Learning Workbench. In: Proc. 2$^{nd}$ Australia and New Zealand Conference on Intelligent Information Systems, Brisbane, Australia (1995)
12. Islam, M. R., Zhou, W., Xiang, Y., Gao, M.: An Innovative Analyser for Multi-Classifier Email Classification Based on Grey List Analysis. In: The Journal of Network and Computer Applications, Elsevier, Vol. 32, No. 2, pp. 357-366 (2009)
13. Islam, M. R., Zhou, W., Xiang, Y., M. Gao, Mahmood, A. N.: Spam Filtering for Network Traffic Security on a Multi-Core Environment. In: Concurrency and Computation: Practice and Experience, Vol. 21, No. 10, pp. 1307-1320 (2009)
14. Ranawana, R., Palade, V.: Multi-Classifier Systems - Review and a Roadmap for Developers. In: International Journal of Hybrid Intelligent Systems, Vol. 3, No. 1, pp. 35-61 (2006)